\documentclass[graybox]{svmult}

\usepackage{mathptmx}       
\usepackage{helvet}         
\usepackage{courier}        
\usepackage{type1cm}        
\usepackage{makeidx}         
\usepackage{graphicx}        
\usepackage{multicol}        
\usepackage[bottom]{footmisc}
\usepackage{mathtools}
\usepackage{amssymb}
\usepackage{float}\usepackage[tight,footnotesize]{subfigure}

\usepackage{todonotes} 

\makeindex             

\graphicspath{{figs/}}
\begin{document}

\title*{A force-based model to reproduce stop-and-go waves in pedestrian dynamics}
\author{M. Chraibi, A. Tordeux and A. Schadschneider}
\institute{
Mohcine Chraibi and Antoine Tordeux
\at Forschungszentrum J\"{u}lich GmbH, J\"{u}lich Supercomputing Centre, J\"{u}lich 52428, Germany,
\email{m.chraibi@fz-juelich.de, a.tordeux@fz-juelich.de}
\and Andreas Schadschneider
\at Institute for Theoretical Physics, Universit\"at zu K\"oln, 
  50937 K\"oln, Germany,
\email{as@thp.uni-koeln.de}}

\maketitle

\abstract*{\texttt{
Stop-and-go waves in single-file movement are a phenomenon that is observed empirically in pedestrian dynamics.
It manifests itself by the co-existence of two phases: moving and stopping pedestrians. 
We show analytically based on a simplified one-dimensional scenario that under some conditions the 
system can have instable homogeneous solutions. 
Hence, oscillations in the trajectories and instabilities emerge during simulations. 
To our knowledge there exists no force-based model which is collision- and oscillation-free and meanwhile can reproduce phase separation.
We develop a new force-based model for pedestrian dynamics able to reproduce qualitatively the phenomenon of phase separation.
 We investigate analytically the stability condition of the model and define regimes of parameter values where phase separation can be observed. 
We show by means of simulations that the predefined conditions lead in fact to the expected behavior and validate our model with respect to empirical findings.
}}

\abstract{
Stop-and-go waves in single-file movement are a phenomenon that is observed empirically in pedestrian dynamics.
It manifests itself by the co-existence of two phases: moving and stopping pedestrians. 
We show analytically based on a simplified one-dimensional scenario that under some conditions the 
system can have instable homogeneous solutions. 
Hence, oscillations in the trajectories and instabilities emerge during simulations. 
To our knowledge there exists no force-based model which is collision- and oscillation-free and meanwhile can reproduce phase separation.
We develop a new force-based model for pedestrian dynamics able to reproduce qualitatively the phenomenon of phase separation.
 We investigate analytically the stability condition of the model and define regimes of parameter values where phase separation can be observed. 
We show by means of simulations that the predefined conditions lead in fact to the expected behavior and validate our model with respect to empirical findings.
}

\section{Introduction}
\label{sec:introduction}

In vehicular traffic, the formation of jams and the dynamics of
traffic waves have been studied intensively~\cite{Chowdhury2000,Orosz2010}. 
Particular car-following models including spacing and speed difference 
variables have been shown to  reproduce realistic stop-and-go phenomena~\cite[chap. 15]{Treiber2013}. 
In pedestrian dynamics this phenomenon has been observed empirically, especially when the density 
exceeds a critical value~\cite{Seyfried2010,Portz2011}. 
Jams can be  reproduced as a result of phase transitions from a
stable homogeneous configuration to an unstable configuration. 
In the literature some space-continuous models
\cite{Portz2010,Seyfried2010b,Lemercier2012,Eilhardt2014} reproduce
partly this phenomenon. However, force-based models generally fail to describe
pedestrian dynamics in jam situations correctly. 
Often uncontrollable oscillations in the direction of motion occur, which lead to
unrealistic dynamics in form of collisions and overlappings~\cite{Chraibi2015}.

In this work we present a force-based model that is able to reproduce 
stop-and-go waves for certain parameter values. By means of a linear stability analysis
we derive conditions to define parameter regions, where the described system 
is unstable.

We study by numerical simulations if the system behaves realistically, i.e.\ 
jams emerge without any collisions in agreement with experimental results~\cite{Portz2011}. 
Furthermore, we validate the model by comparing the fundamental diagram with experiments. 
We conclude this paper with a discussion of the results and the limitations of the
proposed model.

\section{Model definition}
The phenomenon of stop-and-go waves in pedestrian dynamics was investigated experimentally in one-dimensional
scenarios~\cite{Portz2011}. Therefore, we limit our analysis to 1D systems.
Consider $N$ pedestrians distributed uniformly in a narrow corridor 
with closed boundary conditions and neglect the effects of walls on pedestrians. 
Furthermore, for interactions among $N$ pedestrians, we assume
that pedestrian $n$ is only influenced by the pedestrian right in front.

\begin{figure}[h!]
  \centering
  \includegraphics[width=0.75\columnwidth]{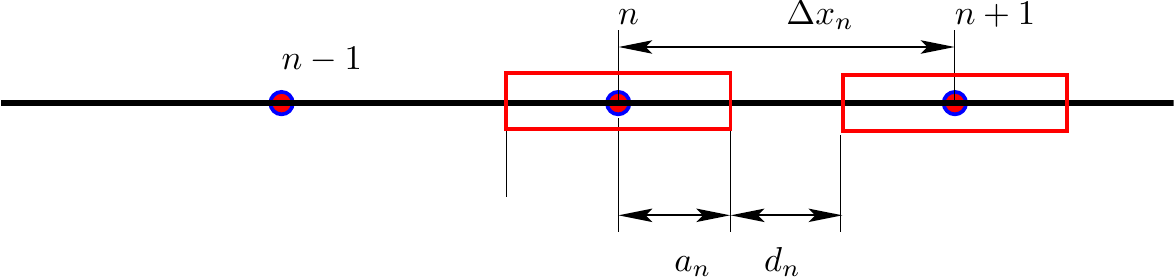}
  \caption{
    Definition of the quantities characterizing the 
    single-file motion of pedestrians (represented by rectangles).}
  \label{fig1}
\end{figure}

For the state variables position $x_n$ and velocity $\dot
x_n=\frac{dx_n}{dt}$ of pedestrian $n$ we define the distance of the
centers $\Delta x_n$ and speed difference $\Delta\dot x_n$ of two
successive pedestrians as
\begin{equation}
  \Delta  x_n = x_{n+1} -  x_n,\qquad
  \Delta\dot x_n = \dot x_{n+1} -  \dot x_n\,.
\end{equation} 
In general, pedestrians are modeled as simple geometric objects of
constant size, e.g.\ a circle or ellipse.  In one-dimensional space
the size of pedestrians is characterized by $a_n$ (Fig.~\ref{fig1}),
i.e.\ their length is $2a_n$. 
However, it is well-known that the space
requirement of a pedestrian depends on its velocity and can be 
characterized by a linear function of the velocity~\cite{Weidmann1993,Jelic2012,Seitz2012}
\begin{equation}
  a_n = a_0+a_v \dot x_n\,,
  \label{eq:an}
\end{equation}
with 
$a_0$, characterizing the space requirement of a standing person
and  $a_v\ge0$ a parameter for the speed dependence with the dimension of time.
The effective distance (distance gap) $d_n$ of two consecutive
pedestrians is then
\begin{equation}                         
 d_n= \Delta x_n - a_n-a_{n+1}=\Delta x_n- a_v\left(\dot x_n +\dot
   x_{n+1}\right)  - 2a_0.                      
 \label{eq:effDist}           
\end{equation}
At each time the change of state variables of pedestrian $n$ is 
given by superposition of driving and repulsive terms. Thus, in general
the equation of motion for pedestrian $n$ described by a force-based model
is given by
\begin{equation}
  \ddot x_n = f\Big(\dot x_n, \Delta \dot x_n, \Delta
  x_n\Big) +  \frac{v_0-\dot x_n}{\tau}.
  \label{eq1}
\end{equation}
Typical values for the parameters are $\tau=0.5$~s for the relaxation
time and $v_0=1.2$~m/s for the desired speed. 

For $f$ we propose the following expression
\begin{equation}
  f(\Delta x_n, \dot x_n, \dot x_{n+1}) = - \frac{v_0}{\tau}
  \ln\Big(c\cdot R_n + 1\Big),
  \label{eq:newf}
\end{equation}
with 
\begin{equation}
  \qquad R_n =r_\varepsilon 
  \Big(\frac{\Delta x_n}{a_n+a_{n+1}} -1\Big),\qquad c = e - 1.
\label{eq:Rn}
\end{equation}
$r_\varepsilon(x)$ is an approximation of the non-differentiable ramp function
\begin{equation}
  r_\varepsilon(x)=\varepsilon \ln(1+e^{-x/\varepsilon})
  \qquad
  (0 < \varepsilon \ll 1).
  \label{eq:aTheta}
\end{equation}

\begin{figure}[H]
  \begin{center}
    \includegraphics[width=0.52\columnwidth]{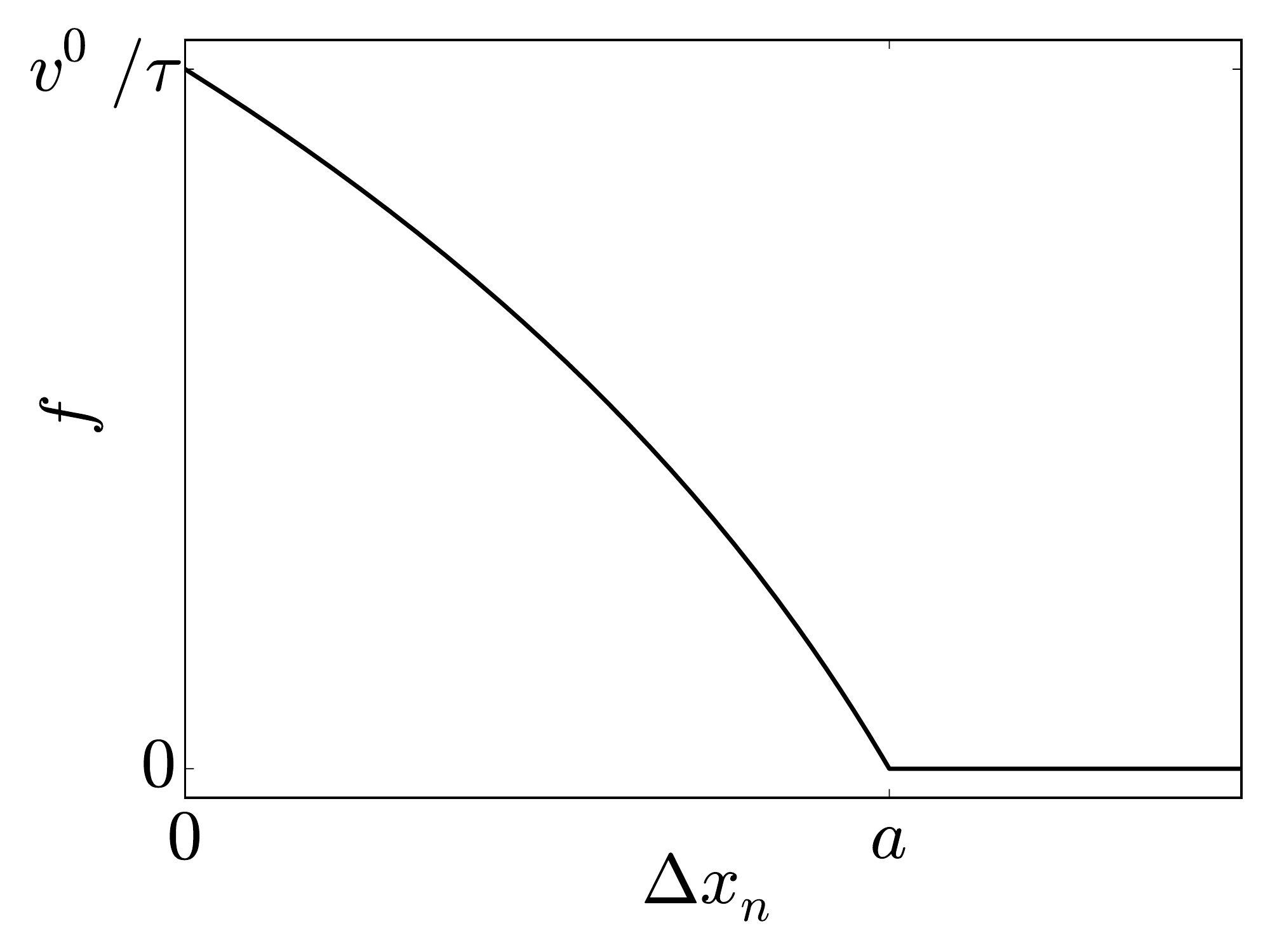}
\vspace{-0.4cm}
    \caption{The absolute value of the repulsive force according to Eq. (\ref{eq:newf}) (at the limit $\varepsilon\rightarrow0$).}
    \label{fig:log_f}
  \end{center}
\end{figure}

Pedestrians anticipate collisions when their distance to their
predecessors is smaller than a critical distance $a=a_n + a_{n+1}$.
Therefore,  $a_n$ does
not only model the body of pedestrian $n$ but represents also a ``personal''
safety distance.  
Assuming that $\dot x_n = 0$, for $\Delta x_n=0$, i.e., $R_n=1$, the repulsive
force reaches the value $-v_0/\tau$ (at the limit $\varepsilon\rightarrow0$) to nullify the effects of the
driving term (Fig.~\ref{fig:log_f}).
\section{Linear dynamics}
In this section, we investigate the stability of the system (\ref{eq1}).  
The position of pedestrian $n$ in the homogeneous steady state is given by
\begin{equation}
  x_n=\frac{n}{\rho} + vt\,, 
\end{equation}
so that
$x_{n+1}-x_n=\frac1{\rho}=\Delta x$, $\dot x_n=v$, 
being speed for the equilibrium of uniform solution.
$\ddot x_n=0$
for all $n$, where derivatives are taken with respect to $t$.
 For $\Delta y = \Delta x_n/a_0$
we consider small (dimensionless) perturbations $\epsilon_n$ 
of the steady state positions of the form
\begin{equation}
  \epsilon_n(t)=\alpha_ne^{z t},
\end{equation}
with $\alpha_n,z\in\mathbb{C}$. 
Replacing in (\ref{eq1}) and expanding to first order yields a second-order equation for $z$. 
To obtain stability, one needs to ensure $\Re(z) < 0$ for the real part of all solutions $z$ with the exception
of the solution $z=0$.

For the system~(\ref{eq1}) with the repulsive force~(\ref{eq:newf}) we obtain the following stability condition
\begin{equation}
  \Phi \coloneqq  \Big(\frac{1}{1 + 2\xi a'_v \Delta
    y}\Big)\Big(\frac{\xi}{1 + 2\xi a'_v \Delta y}  +\xi a'_v\Delta y
  \Big) - 1/2 < 0,
  \label{eq:log_condition}
\end{equation}
with $\xi= \frac{c}{d_0}\frac{v'_0}{a'}, a'=\frac{a}{a_0},\, v'_0=v_0\frac{\tau}{a_0}\;  {\rm{and}}\; d_0 = 1 + c(1-\Delta y/a')$. 

\begin{figure}[H]
  \begin{center}
    \includegraphics[width=0.75\columnwidth]{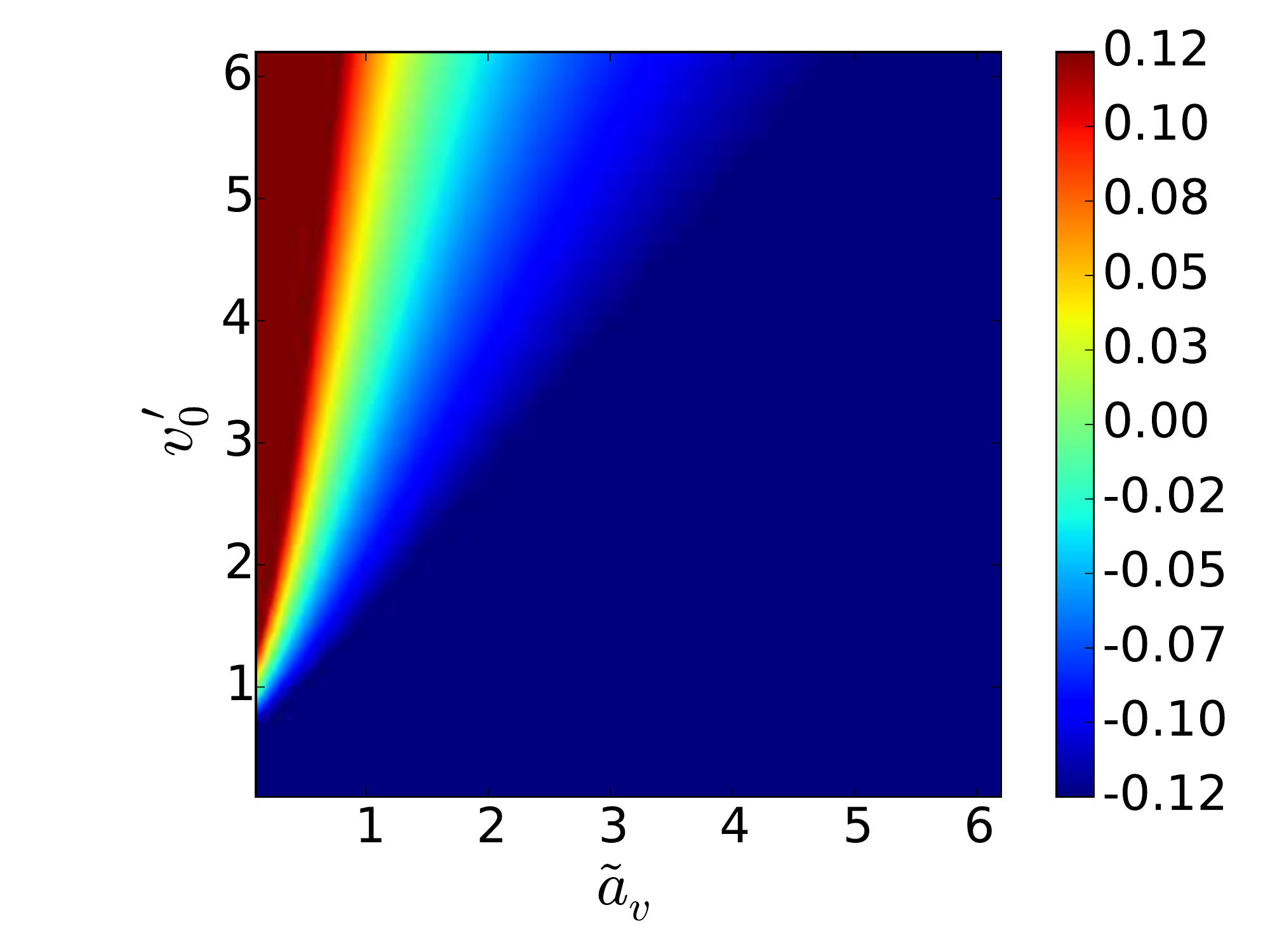}
    \caption{Stability region in the $(\tilde a_v, v^\prime_0)$-space for
      $\Delta y=1.5$.  The colors are mapped to the values of $
      \Phi$ and $(\tilde a_v, v^\prime_0)$ are the dimensionless parameters in Eq.~(\ref{eq:log_condition}).}
    \label{fig:new_av_v0}
  \end{center}
\end{figure}

Fig.~\ref{fig:new_av_v0} shows the stability behavior of the system
with respect to the dimensionless parameters  $v'_0$ and $\tilde a_v= a_v/\tau$. 
The system becomes increasingly unstable with increasing $v'_0$ (for a relatively small
and constant $\tilde a_v$). 
Assuming that the free flow speed $v_0$ is
constant, this means that increasing the reaction
time $\tau$ or diminishing the safety space leads to unstable behavior
of the system. This results is well-known in traffic theory (see for instance~\cite{Bando1995}).

\section{Simulations}
\begin{figure}[H]
  \centering
  \includegraphics[scale=0.35]{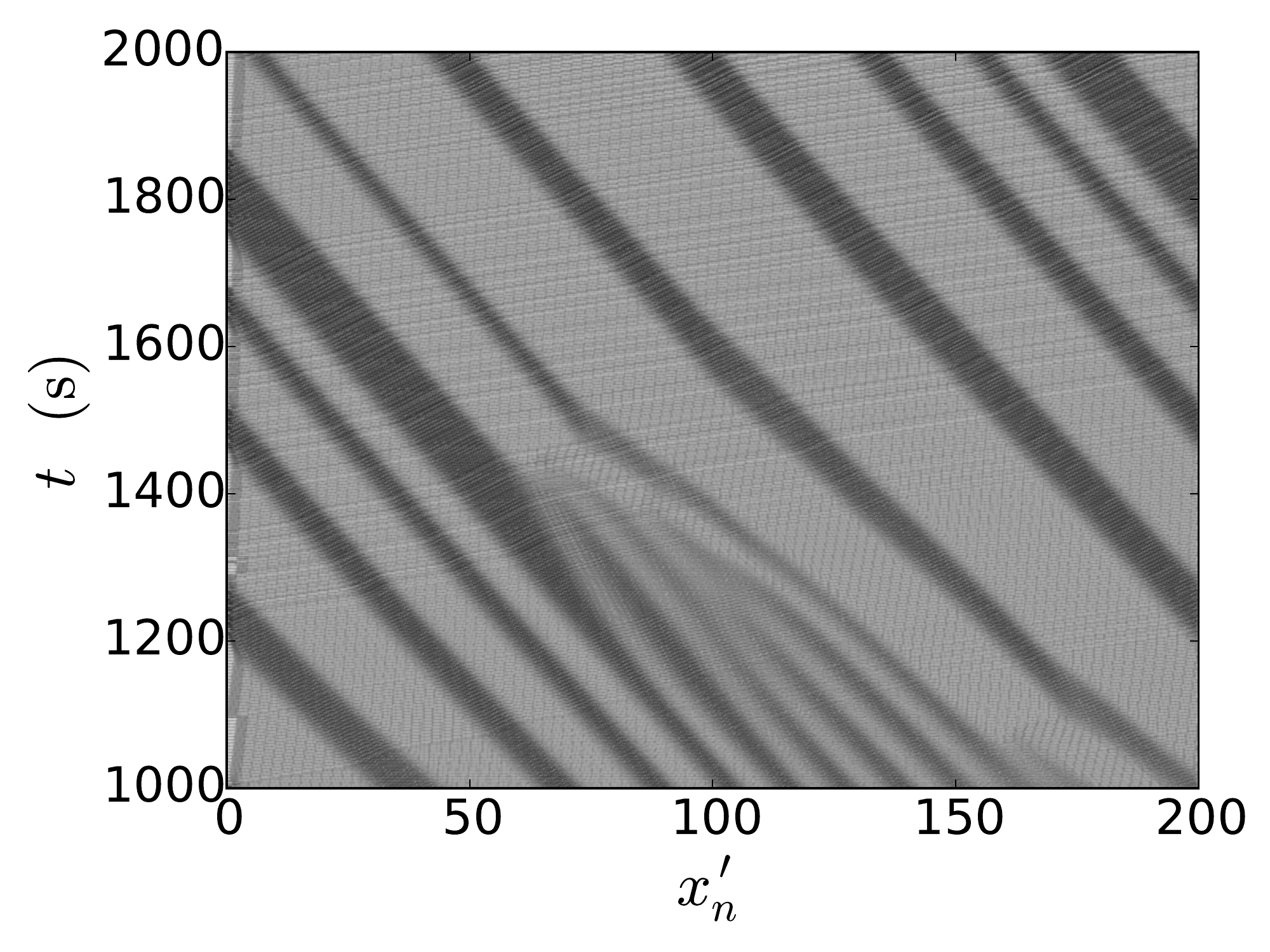}
\vspace{-0.3cm}
  \caption{Trajectories for $\Delta y_n = 1.5$ show stop-and-go waves.}
  \label{fig:jams}
\end{figure}

We perform simulations with the introduced model to analyze the unstable dynamics.
For $a_v=0$, $v'_0=1$  and $\Delta
y_n = 1.5$ we calculate the solution for 3000 s. 
These parameters lay in the unstable regime of the model (Fig.~\ref{fig:new_av_v0}). 
Thus, jam waves are expected to emerge.
Fig.~\ref{fig:jams} shows the trajectories of 133 pedestrians. 
$\varepsilon$ in Eq. (\ref{eq:aTheta}) is set to 0.01.

We observe jam waves propagating in the system. 
Note that the observed jam waves last for a long period of time (here 3000 s), which 
is a indication that they are not dependent on the initial conditions of the simulation and are ``stable''
in time.

\begin{figure}[H]
  \centering
  \subfigure{}
  \includegraphics[scale=0.28]{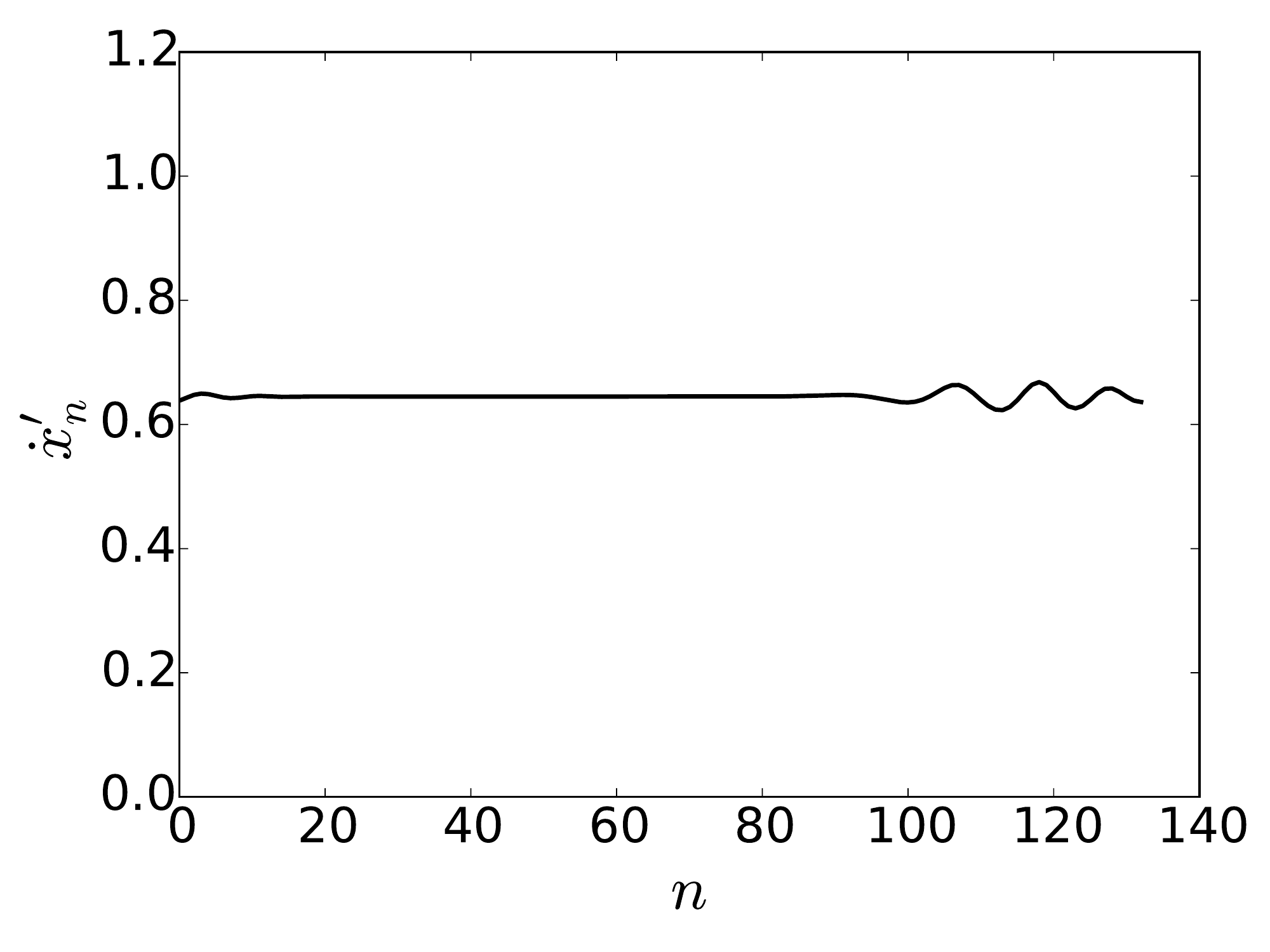}
  \subfigure{}
  \includegraphics[scale=0.28]{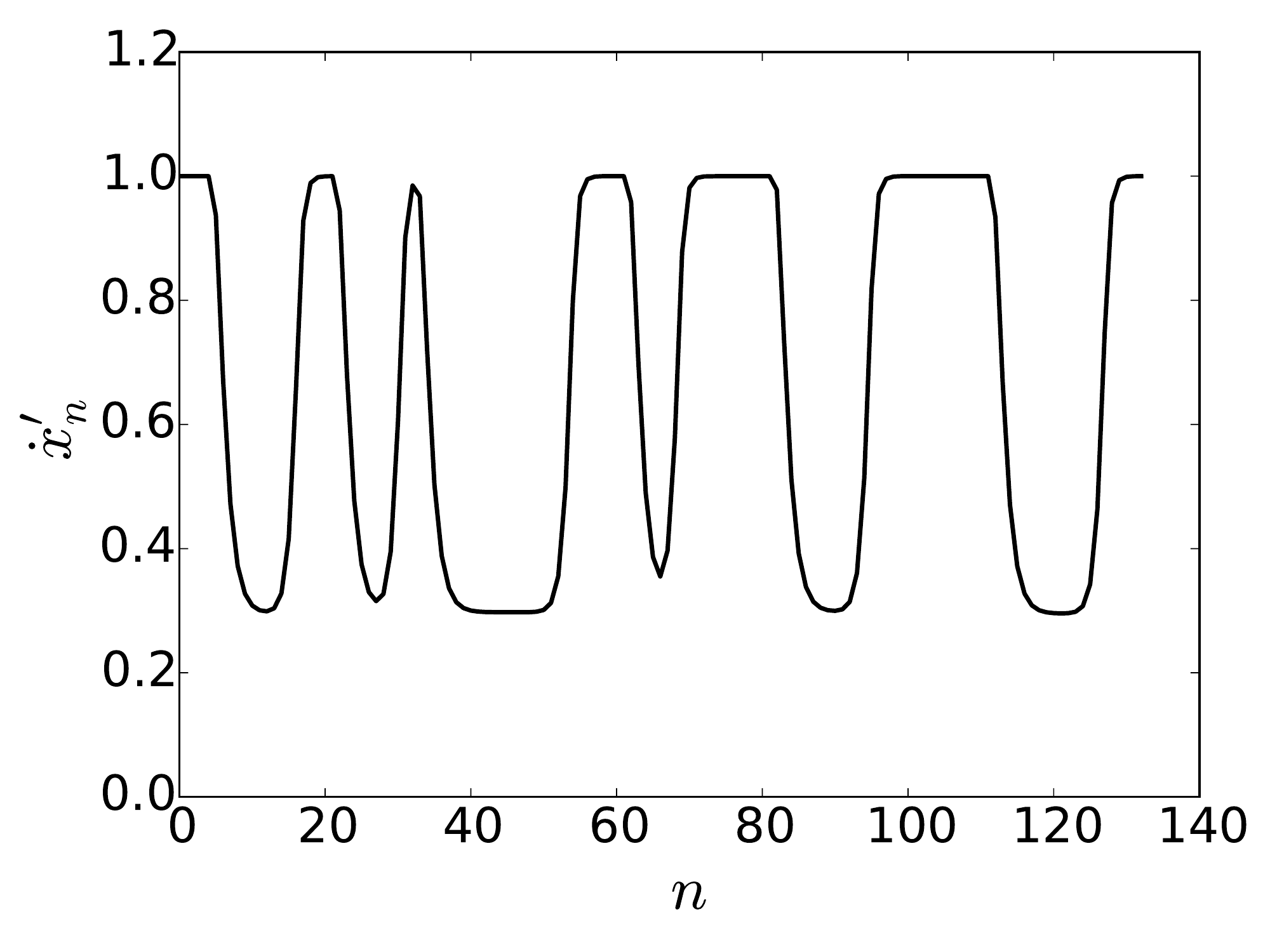}
\vspace{-0.4cm}
  \caption{Speed of pedestrians at different time steps. 
    Left: $t=300\,$ s,
    right: $t=2000\,$ s.}
  \label{fig:jams_vel}
\end{figure}

As shown in Fig.~\ref{fig:jams_vel} the speed does not become negative,
therefore backward movement is not observed.  This condition favors
the appearance of stable jams.

\begin{figure}[H]
  \centering
  \includegraphics[scale=0.45]{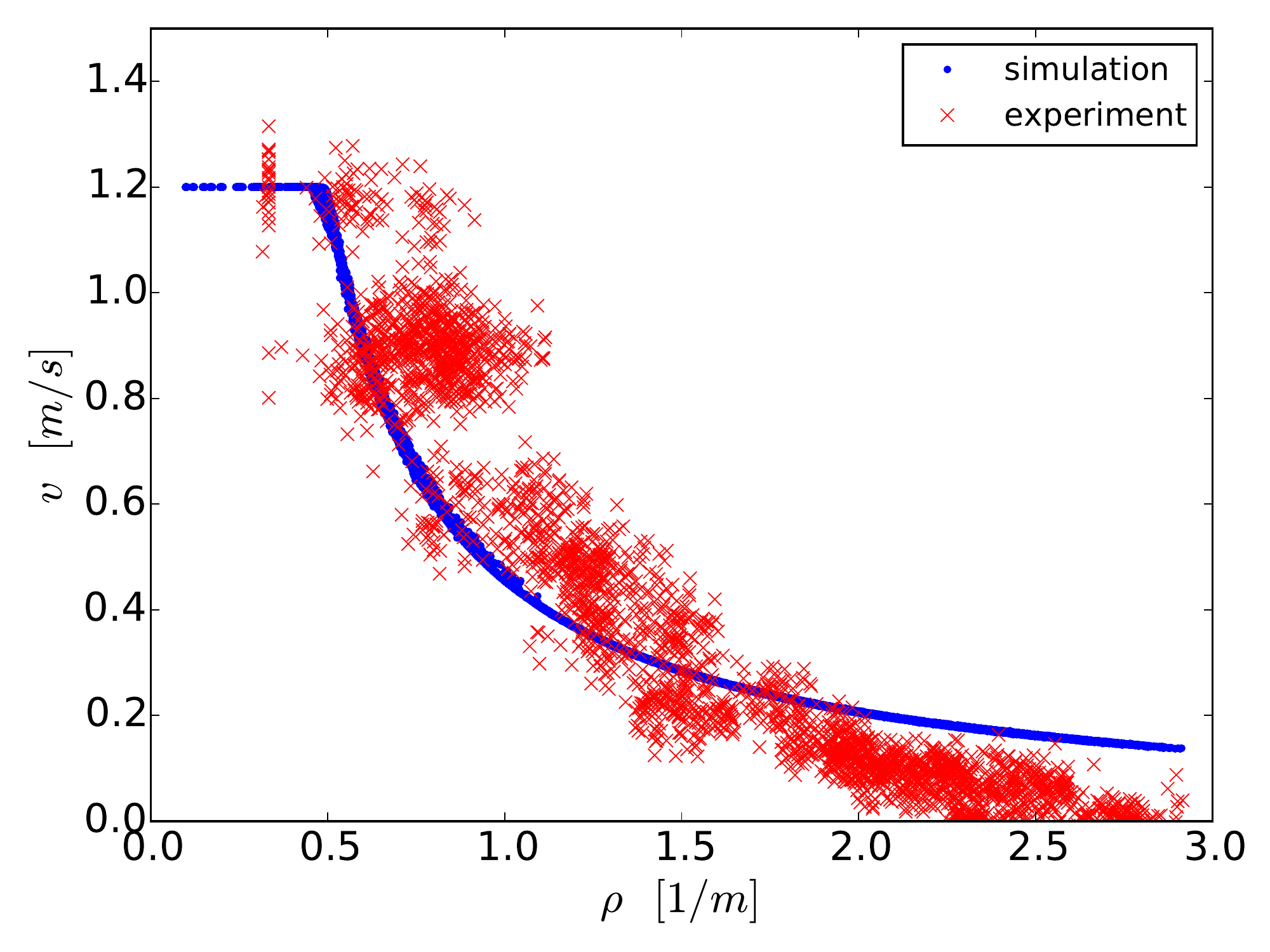}
\vspace{-0.3cm}
  \caption{Fundamental diagram: comparison with experiments from~\cite{Seyfried2010}.}
  \label{fig:fd}
\end{figure}

Having reproduced stop-and-go waves, the model will be further tested by comparing qualitatively the
produced density-velocity relation (fundamental diagram). 
The same setup as above is simulated several times. 
In order to scan a sufficiently large density interval, the number of pedestrians $N$ is increased after each simulation. Fig.~\ref{fig:fd} shows a comparison of the simulation results with experimental data from~\cite{Seyfried2010}. 
The observed fundamental diagram is composed of two different regimes: free flow regime, where the speed of pedestrians does not depend on the density ($\rho < 0.5\, {\rm m}^{-1}$), and a regime where 
the speed decreases with increasing density. 
Here we observe that the correct shape of the fundamental diagram is reproduced quite well, although the velocity is slightly higher than the experimental velocities for $\rho>2 \, {\rm m}^{-1}$.  

\section{Discussion}
We have introduced a simple force-based
model for which uniform solutions can be unstable. 
By simulations we observe that the proposed model shows phase separation in its
unstable regime, in agreement with empirical results~\cite{Portz2011}. 

The linear stability condition of the models shows that we can find \textit{realistic}
parameter values in the unstable regime. 
However, depending on the
chosen values for the (rescaled) desired speed $v'_0$, collisions \textit{can} occur, as a result of backwards movement
and negative speeds. 

Further investigations remain to be carried out to determine the set of parameter 
values for which the model have unstable solutions with 
realistic (i.e.\ collision-free) stop-and-go phenomena and meanwhile a better
\textit{quantitative} agreement with the experimental data e.g.~in form of the fundamental diagram.

\begin{acknowledgement}
   In memory of Matthias Craesmeyer.

  M.C. thanks the Federal Ministry of Education and Research (BMBF) for partly supporting this work under the grant number 13N12045.
  A.S. thanks the Deutsche Forschungsgemeinschaft (DFG) for support
  under grant ``Scha 636/9-1''.
\end{acknowledgement}


\bibliographystyle{spmpsci}
\bibliography{ped2}


\end{document}